\documentclass[twocolumn]{revtex4}
\def\eps{\epsilon}

\def\3nab{\tilde{\nabla}}

\def\hs {\,-\,}

\def\be {\begin{equation}}
\def\ee {\end{equation}}
\def\bea {\begin{eqnarray}}
\def\eea {\end{eqnarray}}

\usepackage{graphicx}
\usepackage{dcolumn}
\usepackage{bm}
\usepackage{longtable}

\begin{document}

\title{Braneworld Dynamics of Inflationary Cosmologies with 
Exponential Potentials}
\author{Naureen Goheer$^1$ and  Peter K.\ S.\ Dunsby$^{1,2}$}

\affiliation{1. Department of Mathematics and Applied Mathematics,
  University of Cape Town, 7701 Rondebosch, Cape Town, South Africa}

\affiliation{2. South African Astronomical Observatory, 
  Observatory 7925, Cape Town, South Africa.}
\begin{abstract}
In this work we consider Randall\hs Sundrum brane\hs world type scenarios, in 
which the spacetime is described by a five\hs dimensional manifold with matter 
fields confined in a domain wall or three-brane.  We present the results of a 
systematic analysis, using dynamical systems techniques, of the qualitative 
behaviour of Friedmann\hs Lema\^{\i}tre\hs Robertson\hs Walker type models, 
whose matter is described by a scalar field with an exponential potential.  
We construct the state spaces for these models and discuss how 
their structure changes with respect to the general\hs relativistic 
case, in particular, what new critical points appear and their 
nature and the occurrence of bifurcation.  
\end{abstract}
\maketitle
\section{Introduction}
String and membrane theories are considered to be 
promising candidates for a unified theory of all forces 
and particles in Nature. Since a consistent construction of 
a quantum string theory is only possible in more than four 
spacetime dimensions we are compelled to compactify any extra 
spatial dimensions to a finite size or, alternatively, find a 
mechanism to localize matter fields and gravity in a lower 
dimensional submanifold.

Motivated by orbifold compactification of higher dimensional string 
theories in particular the dimensional reduction of eleven\hs dimensional
supergravity introduced by Ho$\check{\mbox{r}}$ava and Witten 
\cite{Horava1,Horava2}, Randall and Sundrum showed that for 
non\hs factorisable geometries in five dimensions there exists a 
single massless bound state confined in a domain wall or three\hs 
brane \cite{Randall}. This bound state is the zero mode of the 
Kaluza\hs Klein dimensional reduction and corresponds to the 
four\hs dimensional graviton. This scenario may be described by a 
model consisting of a five\hs dimensional Anti\hs de Sitter space 
(known as the {\it bulk}) with an embedded three\hs brane on which 
matter fields are confined and Newtonian gravity is effectively 
reproduced on large\hs scales. A discussion of earlier work on Kaluza\hs 
Klein dimensional reduction and matter localization in a four\hs 
dimensional manifold of a higher\hs dimensional non\hs compact spacetime 
can be found in \cite{List1}.

Gravity on the brane can be described by the standard Einstein equations
modified by two additional terms, one quadratic in the energy\hs momentum
tensor and the other representing the electric part of the five\hs dimensional
Weyl tensor.

Following an approach to the study of homogeneous cosmological models with a
cosmological constant first introduced by by Goliath and Ellis \cite{Goliath}
(see also \cite{Dynamics} for a detailed discussion of the dynamical systems 
approach to cosmology), Campus and Sopuerta have recently studied the 
complete dynamics of Friedmann\hs Lema\^{\i}tre\hs Robertson\hs 
Walker (FLRW) and the Bianchi type I and V cosmological models 
with a barotropic equation of state, taking into account effects 
produced by the corrections to the Einstein Field Equations \cite{Campos}.  

Their analysis led to the discovery of new critical points corresponding 
to the Bin$\acute{\mbox{e}}$truy\hs Deffayet\hs Langlois (BDL) models 
\cite{Binetruy}, representing the dynamics at very high energies, 
where effects due to the extra dimension become dominant. These solutions
appear to be a generic feature of the state space of more general 
cosmological models. They also showed that the state space contains new 
bifurcations, demonstrating how the dynamical character of some of the 
critical points changes relative to the general\hs relativistic case.
Finally, they  showed that for models satisfying all the ordinary energy 
conditions and causality requirements the anisotropy is negligible near 
the initial singularity, a result first demonstrated by Maartens 
{\it et. al.} \cite{Maartens-Sahni}.

In this paper we extend their work to the case where the matter is described 
by a dynamical scalar field $\phi$ with an exponential potential 
$V(\phi)=\exp(b\phi)$. This work builds on earlier results obtained by 
Burd and Barrow \cite{Burd} and Haliwell \cite{Halliwell} who considered 
the dynamics of these models in standard general relativity (GR). 
Van den Hoogen {\it et. al.} and 
Copeland {\it et. al.} have also considered the GR dynamics of exponential
potentials in the context of {\it Scaling Solutions} in FLRW spacetimes
containing an additional barotropic perfect fluid \cite{Hoogen,Copeland}.

It is worth mentioning that Exponential potentials are somewhat more 
interesting in the brane\hs world scenario, since the high\hs 
energy corrections to the Friedmann equation allow for inflation to 
take place with potentials ordinarily too steep to sustain inflation 
\cite{Copeland2}.

\section{Preliminaries}\label{Preliminaries}
In what follows we summarise the geometric framework used analyse
the brane\hs world scenario in the cosmological context.
\subsection{Basic equations of the brane\hs world}
In Randall\hs Sundrum brane\hs world type scenarios matter fields are
confined in a three\hs brane embedded in a five\hs dimensional 
spacetime (bulk). It is assumed that the metric of this spacetime,
$g^{(5)}_{AB}$, obeys the Einstein equations with a negative 
cosmological constant $\Lambda_{(5)}$ \cite{Shiromizu,Sasaki,Maartens})
\begin{equation}
G^{(5)}_{AB} = -\Lambda_{(5)}g^{(5)}_{AB}
+\kappa^2_{(5)} \delta(\chi)\left[ -\lambda\,g_{AB}+T_{AB}
\right]\;,  \label{Einstein}
\end{equation}
where $G^{(5)}_{AB}$ is the Einstein tensor, $\kappa_{(5)}$ denotes 
the five\hs dimensional gravitational coupling constant and  
$T_{AB}$ represents the energy\hs momentum tensor of the matter with the 
Dirac delta function reflecting the fact that matter is confined 
to the spacelike hypersurface $x^4\equiv\chi=0$ (the brane) with induced 
metric $g_{AB}$ and tension $\lambda$.   

Using the Gauss\hs Codacci equations, the Israel junction conditions and
the $Z_2$ symmetry with respect to the brane the effective  
Einstein equations on the brane are
\begin{equation}
G_{ab}=-\Lambda g_{ab}+\kappa^2 T_{ab}+
\kappa^4_{(5)} S_{ab} - {\cal E}_{ab} \,, \label{Modified}
\end{equation}
where $G_{ab}$ is the Einstein tensor of the induced metric $g_{ab}$. 
The four\hs dimensional gravitational constant $\kappa$ and the 
cosmological constant $\Lambda$ can be expressed in terms of the 
fundamental constants in the bulk $(\kappa_{(5)},\Lambda_{(5)})$ and 
the brane tension $\lambda $ \footnote{In order to recover conventional 
gravity on the brane $\lambda$ must be assumed to be positive.}.

As mentioned in the introduction, there are two corrections to the 
general\hs relativistic equations. Firstly $S_{ab}$ represent corrections 
quadratic in the matter variables due to the form of the Gauss\hs Codacci 
equations:
\begin{equation}
S_{ab} = \textstyle{\frac{1}{12}}T T_{ab}-\textstyle{\frac{1}{4}}
T_a{}^c T_{bc}+\textstyle{\frac{1}{24}}g_{ab}\left[3T^{cd}
T_{cd}-T^2\right] \;. \label{Sab}
\end{equation}
Secondly ${\cal E}_{ab}$, corresponds to the ``electric'' part of the 
five\hs dimensional Weyl tensor $C^{(5)}_{ABCD}$ with respect to the 
normals, $n_A$ ($n^An_A=1$), to the hypersurface $\chi=0$, that is
\begin{equation}
{\cal E}_{AB} = C^{(5)}_{ACBD}n^Cn^D \;,
\end{equation}
representing the non\hs local effects from the free gravitational 
field in the bulk.
The modified Einstein equations (\ref{Modified}) together with the 
conservation of energy\hs momentum equations $\nabla^a T_{ab}=0$ lead
to a constraint on $S_{ab}$ and ${\cal E}_{ab}$:
\begin{equation} 
\nabla^a ({\cal E}_{ab}-\kappa^4_{(5)}S_{ab}) = 0 \,. \label{Conserved}
\end{equation}
We can decompose ${\cal E}_{ab}$ into its ineducable parts relative 
to any timelike observers with 4\hs velocity $u^a$ ($u^au_a=-1$):
\begin{equation} 
{\cal E}_{ab} = -\left(\frac{\kappa^{}_{(5)}}{\kappa}\right)^4
\left[ (u_au_b+\textstyle{\frac{1}{3}}h_{ab}){\cal U} +2u_{(a}{\cal Q}_{b)}
+{\cal P}_{ab}\right]\,, \label{spli}
\end{equation}
where 
\begin{equation}
{\cal Q}_au^a=0\,,~~~ {\cal P}_{(ab)}= {\cal P}_{ab}\,,~~
{\cal P}^a{}_a=0\,,~~ {\cal P}_{ab}u^b=0 \,. 
\end{equation}
Here ${\cal U}$ has the same form as the energy\hs momentum tensor 
of a radiation perfect fluid and for this reason is referred to
as the ``dark'' energy density of the Weyl fluid. ${\cal Q}_a$ is a 
spatial and ${\cal P}_{ab}$ is a spatial, symmetric and trace\hs 
free tensor.  ${\cal Q}_a$ and ${\cal P}_{ab}$ are analogous to the usual 
energy flux vector $q^a$ and anisotropic stress tensor $\pi_{ab}$ in 
General Relativity. The constraint equation (\ref{Conserved}) leads to 
evolution equations for ${\cal U}$ and ${\cal Q}_a$, but not 
for ${\cal P}_{ab}$ (see \cite{Maartens}). 

The above equations correspond to the general situation. In what 
follows we restrict our analysis to the case where ${\cal E}_{ab}=0$
so the brane is conformally flat. Such bulk spacetimes admit
FLRW brane\hs world models with vanishing non\hs local energy density
(${\cal U}=0$). 
\subsection{Scalar field dynamics on the brane}
Relative to a normal congruence of curves with tangent vector
\begin{equation}
u^a=-\frac{\nabla_a\phi}{\dot{\phi}}\;,~~u^au_a=-1\;,
\label{Four-velocity}
\end{equation} 
the energy\hs momentum tensor $T_{\mu\nu}$ for a scalar field 
takes the form of a perfect fluid (See page 17 in \cite{BED} for details):
\begin{equation}
T_{\mu\nu}=\rho u_{\mu}u_{\nu}+ph_{\mu\nu}\,,\label{Scalar-EMT}
\end{equation}
with
\begin{equation}
\rho=\frac{1}{2}\dot{\phi}^2+V(\phi)\label{Scalar-energy}
\end{equation}
and 
\begin{equation}
p=\frac{1}{2}\dot{\phi}^2-V(\phi)\;.\label{Scalar-momentum}
\end{equation}
where $\dot{\phi}$ is the momentum density of the scalar field and 
$V(\phi)$ is its potential energy. If the scalar field is not 
minimally coupled this simple representation is no longer valid, 
but it is still possible to have an imperfect fluid form for the 
energy\hs momentum tensor \cite{Madsen}.

Substituting for $\rho$ and $p$ from (\ref{Scalar-energy}) and 
(\ref{Scalar-momentum}) into the energy conservation equation 
\begin{equation}
\dot{\rho} +\Theta(\rho +p)=0\;,
\label{Energy-conservation} 
\end{equation}
leads to the 1+3 form of the Klein\hs Gordon equation 
\begin{equation}
\ddot{\phi} +\Theta \dot{\phi} +V'(\phi)=0\;, \label{Klein-Gordon}
\end{equation}
an exact ordinary differential equation for $\phi$ once the potential
has been specified.
It is convenient to relate $p$ and $\rho$ by the {\em index $\gamma$} 
defined by 
\begin{equation}
p = ( \gamma - 1) \rho  ~~\Leftrightarrow ~~ \gamma = \frac{p + \rho}{\rho}
~=~\frac{\dot{\phi}^2}{\rho}\;.
\end{equation}
This index would be constant in the case of a simple one\hs component 
fluid, but in general will vary with time in the case of a scalar 
field:
\begin{equation}
\\{\dot{\gamma}}=\Theta\gamma(\gamma-2)-2\gamma\frac{V'}{\dot{\phi}}\;.
\label{Gammadot}
\end{equation}
Notice that this equation is well-defined even for $\dot{\phi}\rightarrow 0$,
since $\frac{\gamma}{\dot{\phi}}=\frac{\dot{\phi}}{\rho}$.

The dynamics of FLRW models imposed by the modified Einstein equations
are governed by the Raychaudhuri and Friedmann equations
\bea
\dot{H}&=&-H^2-\frac{3\gamma-2}{6}\kappa^2\rho\left[1+\frac{3\gamma-1}
{3\gamma-2}\frac{\rho}{\lambda}\right]\;,
\label{Raychaudhuri}\\
H^2&=&\frac{1}{3}\kappa^2\rho\left(1+\frac{\rho}{2\lambda}\right)
-\frac{1}{6}{}^3R\;, \label{Friedmann}
\eea
together with the Klein Gordon equation (\ref{Klein-Gordon}) above. $^3R$
denotes the scalar curvature of the hypersurfaces orthogonal to the 
the fluid 4\hs velocity (\ref{Four-velocity}) and can be expressed
in terms of the scale factor via $^3 R={\textstyle\frac{6k}{a^2}}$ where
as usual $k=0,\pm 1$ determines whether the model is flat, open or closed. 
\section{Dynamical systems analysis for exponential potentials}
In the following analysis we extend recent work presented by Campus 
and Sopuerta \cite{Campos} to the case where the matter is described 
by a dynamical scalar field $\phi$ with a self\hs interacting potential 
$V(\phi)=\exp(b\phi)$, where $b\leq 0$ and consider only models which 
have negligible non\hs local energy density ${\cal U}$ and cosmological 
constant $\Lambda$. Although the full dynamics are described by the
equations presented in the previous section, it is useful 
to re\hs write these equations in terms of a set of dimensionless 
expansion normalized variables. In order to get compact state spaces 
it is convenient to consider two different cases 
(i) $^3 R\leq 0$ ($k=0$ or $k=-1$) and (ii) $^3 R\geq 0$ ($k=0$ or $k=1$). In case 
(i) the appropriate variables are 
\bea
\Omega_{\rho}\equiv\frac{\kappa^2\rho}{3H^2},~~\Omega_{k} \equiv 
-\frac{^3R}{6H^2}=-\frac{k}{\dot{a}^2},~~\Omega_{\lambda} \equiv 
\frac{\kappa^2\rho^2}{6\lambda H^2}\;,
\eea
leading via equation (\ref{Friedmann}) to very simple expression 
of the Friedmann constraint: 
\begin{equation}
\label{friedo}
\Omega_{\rho}+\Omega_{k}+\Omega_{\lambda}=1\;.
\end{equation}
We introduce a dimensionless time variable by 
\begin{equation}
\nonumber
' \equiv \frac{1}{|H|}\frac{d}{dt}\;,
\end{equation} 
where $|H|$ is the absolute value of $H$. It follows that 
\begin{equation}
H'=-\epsilon (1+q)H\;,
\end{equation} 
where $\epsilon$ is the sign of $H$ and $q$ is the usual deceleration 
parameter defined by 
\begin{equation}
q=-\frac{1}{H^2}\frac{\ddot{a}}{a}\;.
\end{equation}
It is clear that $\epsilon=1$ corresponds to models which are expanding 
while $\epsilon=-1$ correspond to models which are contracting.

Using the above definitions, the dynamics for open and flat models are 
described by the following system of equations
\bea
\label{omega_k.o}
\Omega_{k}'=\eps[(3\gamma-2)(1-\Omega_{k})+3\gamma\Omega_{\lambda}]
\Omega_{k}\;,\\
\nonumber
\Omega_{\lambda}'=\eps[3\gamma(\Omega_{\lambda}-\Omega_{k}-1)
+2\Omega_{k}]\Omega_{\lambda}\;.
\eea
In case (ii) it is necessary to normalise using 
\begin{equation}
D \equiv \sqrt{H^2+{\textstyle\frac{1}{6}}{^3R}}
\end{equation}
instead of the Hubble function $H$, so the dimensionless variables for 
this case are given by 
\bea
Q \equiv \frac{H}{D},~~\tilde{\Omega}_{\rho} \equiv \frac{\kappa^2\rho}{3D^2}
\,~~\tilde{\Omega}_{\lambda} \equiv \frac{\kappa^2\rho^2}{6\lambda D^2}
\eea   
and the Friedmann constraint is 
\begin{equation}
\label{friedc}
\tilde{\Omega}_{\rho}+\tilde{\Omega}_{\lambda}=1\;.
\end{equation}
The appropriate dimensionless time derivative is defined via
\begin{equation}
\nonumber
' \equiv \frac{1}{D}\frac{d}{dt}
\end{equation}
so the dynamical equations for closed models become
\bea
\label{Q.}
Q'=[1-\frac{3}{2}\gamma(1+\tilde{\Omega}_{\lambda})](1-Q^2)\;,\\
\nonumber
\tilde{\Omega}_{\lambda}'=3\gamma Q(\tilde{\Omega}_{\lambda}-1)
\tilde{\Omega}_{\lambda}\;.
\eea
Notice that in (\ref{omega_k.o}), (\ref{Q.}) we have already included the 
constraint (\ref{friedo}), 
(\ref{friedc}) in order to keep the dimensionality of the state space 
as low as possible. The evolution of $\Omega_{\rho}$, 
$\tilde{\Omega}_{\rho}$ can easily be recovered from the Friedmann equation.

We have to add a third equation to (\ref{omega_k.o}), (\ref{Q.}) respectively 
in order to describe the dynamics of the scalar field $\phi$. It 
turns out, that the equation of state parameter $\gamma$ instead 
of $\dot{\phi}$, is a the preferred coordinate, since it is both 
bounded by causality requirements ($0 \leq\gamma\leq 2$) and is 
dimensionless.     

For open models (case (i)), we find using (\ref{Gammadot}) that the 
evolution of $\gamma$ is determined by
\begin{equation}
\gamma '=\eps \sqrt{3\gamma} (\gamma -2)[\sqrt{3\gamma}+\eps~sgn(\dot{\phi}) 
b \sqrt{1-\Omega_{k}-\Omega_{\lambda}}]
\label{gamma.o}
\end{equation}
where $sgn(\dot{\phi})$ is the sign of $\dot{\phi}$. 
We can see from (\ref{gamma.o}) that we only need to consider the 
case $\dot{\phi} \geq 0$, since the case $\dot{\phi}\leq 0$ can be 
recovered from the former by time reversal; simultaneously changing the sign 
of $\dot{\phi}$ and $H$ results in an overall change of sign of 
$\gamma'$. From (\ref{omega_k.o}), we see that this transformation 
also changes the sign of $\Omega_{k}'$ and $\Omega_{\lambda}'$, which 
means that $\dot{\phi}\rightarrow -\dot{\phi}$ corresponds to time 
reversal $\tau \rightarrow -\tau$.

In the closed case, we find that
\begin{equation}
\gamma '= \sqrt{3\gamma} (\gamma -2)[\sqrt{3\gamma}Q + sgn(\dot{\phi}) b \sqrt{1-\tilde{\Omega}_{\lambda}}]
\label{gamma.c}
\end{equation}
and again we can easily see from (\ref{gamma.c}) and (\ref{Q.}) 
that $\dot{\phi}\rightarrow -\dot{\phi}$ corresponds to time 
reversal $\tau \rightarrow -\tau$. Therefore in the following analysis we 
restrict ourselves to the case $\dot{\phi} \geq 0$.

So in summary, the resulting dynamical equations we have to analyse are
\bea
\label{system.o}
\gamma '=\eps \sqrt{3\gamma} (\gamma -2)[\sqrt{3\gamma}
+\eps~ b \sqrt{1-\Omega_{k}-\Omega_{\lambda}}]\;,\\
\nonumber
\Omega_{k}'=\eps[(3\gamma-2)(1-\Omega_{k})
+3\gamma\Omega_{\lambda}]\Omega_{k}\;,\\
\nonumber
\Omega_{\lambda}'=\eps[3\gamma(\Omega_{\lambda}-\Omega_{k}-1)
+2\Omega_{k}]\Omega_{\lambda}
\eea
for the open case and
\bea
\label{system.c}
\gamma '= \sqrt{3\gamma} (\gamma -2)[\sqrt{3\gamma}Q + 
b \sqrt{1-\tilde{\Omega}_{\lambda}}]\;,\\
\nonumber
Q'=[1-\frac{3}{2}\gamma(1+\tilde{\Omega}_{\lambda})](1-Q^2)\;,\\
\nonumber
\tilde{\Omega}_{\lambda}'=3\gamma 
Q(\tilde{\Omega}_{\lambda}-1)\tilde{\Omega}_{\lambda}
\eea
for  the closed case.

Notice that in the open sector, the cosmological constant\hs like 
subset $\gamma=0$, the stiff\hs matter\hs subset $\gamma=2$, the flat 
subset $\Omega_{k}=0$, the GR subset $\Omega_{\lambda}=0$ and the 
vacuum subset $\Omega_{k}+\Omega_{\lambda}=1$ are invariant sets. 
Analogously, the closed sector has the invariant sets $\gamma=0 ,2$, 
the flat subset $Q=\pm1$, the GR subset $\tilde{\Omega}_{\lambda}=0$ 
and the vacuum subset $\tilde{\Omega}_{\lambda}=1$.

\begin{table*}
\caption{\label{tab:table1} This table gives the coordinates and 
eigenvalues of the critical points with non\hs positive spatial curvature. 
We have defined $\psi=\sqrt{\frac{8}{b^2}-3}$.}
\begin{ruledtabular}
\begin{tabular}{ccc}
Model  & Coordinates   & Eigenvalues \\ \hline
$\mbox{m}_\epsilon^0(\Omega_{\lambda}\neq 1)~for~ b=0$ & 
$(0,0,\Omega_{\lambda})$ & $-2\epsilon(3,1,0)$ \\
$\mbox{m}_\epsilon^0(\Omega_{\lambda}\neq 1)~for~ b 
\neq 0$ & $(0,0,\Omega_{\lambda})$ & $(\infty,-2\epsilon,0)$ \\
$\mbox{F}_\epsilon^2$ & $(2,0,0)$ & $(6\epsilon+\sqrt{6}b,4\eps,-6\eps)$ \\
$\mbox{M}_\epsilon^2$ & $(2,1,0)$ & $2\epsilon(3,-2,-5)$ \\
$\mbox{m}_\epsilon^2$ & $(2,0,1)$ & $2\epsilon(3,5,3)$ \\
$\mbox{F}_{+}^{b^2/3}$ & $(\frac{b^2}{3},0,0)$ & 
$(\frac{b^2}{2}-3,b^2-2,-b^2)$\\
$\mbox{X}_{+}^{2/3}(b)$ & $(\frac{2}{3},1-\frac{2}{b^2},0)$ 
& $(-1-\psi,-1+\psi,-2)$\\
\end{tabular}
\end{ruledtabular}
\end{table*}
\section{Analysis of the Dynamical System}
In order to analyse the dynamical systems (\ref{system.o}) and 
(\ref{system.c}), we will use for the most part the standard method 
of linearising the dynamical equations around any equilibrium points. 
One can easily see that because of the $\gamma'$\hs equation, 
the linearisation matrix (the Jacobian), is not well\hs defined for 
certain values of $(\gamma, \Omega_k, \Omega_{\lambda})$ if $b\neq0$.
Indeed, if $b\neq0$ and $\gamma=0$ or $b\neq0$ and 
$\Omega_k+\Omega_{\lambda}=1$ ($\tilde{\Omega}_{\lambda} =1$), the 
Jacobian diverges. This means that we cannot linearise the dynamical 
equations in the neighbourhood of these points. Instead, we will have 
to consider the full non\hs linear equations and study the behaviour 
of small perturbations away from these problematic equilibrium points.

Furthermore the fact that we are dealing with a non\hs linear system is 
also important in cases, where the dynamical equations can be linearised. 
Whenever the linear terms are not dominant, the peculiar behaviour of 
non\hs linear systems becomes visible. We then have to analyse the 
dynamical equations with additional caution. If for example an 
eigenvalue of the linearised system vanishes at an equilibrium point, 
which means that the first order terms of the dynamical equation vanishes, 
we have to study the higher order terms in a perturbative analysis around 
that point.

Notice that the dynamical systems (\ref{system.o}) and (\ref{system.c}) 
match at the $k=0$ plane (the flat subspace). If we analyse critical 
points that lie in the flat subspace by studying small perturbations 
out of that surface, i.e. where (\ref{system.o}) differs from 
(\ref{system.c}), it is necessary to do the analysis in both coordinate 
systems and check, whether the results agree. If they don't, this means 
that small perturbations around the critical point will evolve 
differently, depending on whether they enter the closed or the 
open sector. In our analysis we have checked this carefully. In only 
one case ($F^{2/3}_+$) the higher order terms were dominant and of 
different sign, depending on which sector was entered. This will be 
commented on below. In all the other cases, the behaviour of the 
perturbations did not depend on the sector they entered \footnote{Note 
however that even if the eigenvalues are the same in the closed and 
open sectors, the eigenvectors pointing out of the $\Omega_k=0$ 
plane into these two different sectors are not necessarily parallel.}. 
\subsection{Models with Non\hs positive Spatial Curvature} \label{Positive}
The dynamical system (\ref{system.o}) has 5 hyperbolic critical
points corresponding to the flat FLRW universe $F_{+}^{b^2/3}$ with
$\gamma=\frac{b^2}{3}$ and $a(t)=t^{2/b^2}$ \footnote{If $b^2=6,~\eps=1$, 
this point is actually non\hs hyperbolic. That case will be discussed in 
detail below.}; the flat FLRW universe  $F_{\epsilon}^2$  with stiff 
matter and $a(t)=t^{1/3}$; the Milne universe $M_{\epsilon}^2$ with 
stiff matter and  $a(t)=t$; a flat non\hs general\hs relativistic model 
$m_{\epsilon}^{2}$ with $\gamma=2$, which has been discussed 
in \cite{Campos}; and a set of universe models $X_{+}^{2/3}(b)$ 
with $\gamma=2/3$ and curvature $\Omega_{k}=1-2/b^{2}$ depending on 
the value of b. The critical points, their coordinates in 
state space and their eigenvalues is given in Table I below. In Table II we  
give the corresponding eigenvectors. Notice 
that $F_{+}^{b^2/3}$ only occurs for $0\leq b^{2}\leq 6$, and 
$X_{+}^{2/3}(b)$ for $b^{2}\geq 2$ only. Both only occur in the 
expanding sector $\epsilon=1$.

We further find the non\hs hyperbolic critical points $M_{\epsilon}^0$ and
$m_{\epsilon}^0(\Omega_{\lambda})$, where the former describes the 
Milne universe with $\gamma=0$ and the latter is a line of non\hs general\hs 
relativistic critical points with $\gamma=0$ including the flat 
FLRW model with constant energy density. The Jacobian of the dynamical 
system (\ref{system.o}) at both $M_{\epsilon}^0$ and 
$m_{\epsilon}^0(\Omega_{\lambda})$ diverges for $b\neq 0$, which means 
that the dynamical system cannot be linearised around these points 
for arbitrary positive values of $b^2$.
 
For $b=0$, the Jacobian is well\hs defined for both
$M_{\epsilon}^0$ and $m_{\epsilon}^0(\Omega_{\lambda})$, and the
eigenvalues and corresponding eigenvectors are given in Tables I, II. 
For $ \Omega_{\lambda}\neq1$, the Jacobian around 
$m_{\epsilon}^0(\Omega_{\lambda})$ can still be evaluated even for 
$b\neq0$. The results from taking the limits ${\gamma , 
\Omega_{k} \rightarrow 0}$ are included in Table I.

If $ \Omega_{\lambda}=1$ and $b^2>0$, we have to
look at the non\hs linear system (\ref{system.o}) and study small
perturbations $x (\tau) , y (\tau) , z (\tau)$ about the equilibrium 
point $m_{\epsilon}^0(\Omega_{\lambda}=1)$. That means 
we evaluate (\ref{system.o}) at $( \gamma , \Omega_{k} ,
\Omega_{\lambda} )=( x , y , 1-z )$, where $0 < x
(\tau) , y(\tau) , z(\tau) \ll 1$.

Up to first order, (\ref{system.o}) becomes
\bea
\nonumber
x'=-6 \epsilon x - 2 \sqrt{3} b \sqrt{x} \sqrt{z-y)}\;,\\
\nonumber
y'=-2 \epsilon y\;,\\
\nonumber
z'=-2 \epsilon y\;.
\eea
For $y=z$ or $b=0$, this can be solved to give
\bea
\nonumber
x = x_{0} e^{-6 \epsilon \tau}\;,\\
\nonumber
y = y_{0} e^{-2 \epsilon \tau}\;,\\
\nonumber
z = z_{0}+y_{0}e^{-2 \epsilon \tau}\;.
\eea
For $y \neq z$ and $b\neq 0$, we find
\bea
\nonumber
x = 3 b^{2} z_{0} (\tau - \tau_{0})^2\;,\\
\nonumber
y = y_{0} e^{-2 \epsilon \tau}\;,\\
\nonumber
z = z_{0}+y_{0}e^{-2 \epsilon \tau}\;,
\eea
where $x_{0}$, $y_{0}$ and $z_{0}$ are positive constants of integration 
(initial values).

This demonstrates, that for $b\neq 0$, $m_{+}^0(\Omega_{\lambda}=1)$ is an
unstable saddle point (to be precise, it is a line of saddles, 
since $z$ is stationary), whereas the contracting model 
$m_{-}^0(\Omega_{\lambda}=1)$ remains a source (actually a line 
of sources) for all b.

In the same way, we analyse the nature of the point
$M_{\eps}^0$. Here, we find that the perturbed
system $( \gamma , \Omega_{k} , \Omega_{\lambda} )=( x
, 1-y , z )$ becomes up to first order
\bea
\nonumber
x'=-6 \epsilon x-2 \sqrt{3} b \sqrt{x} \sqrt{y-z}\;,\\
\nonumber
y'=2 \epsilon y\;,\\
\nonumber
z'=2 \epsilon z\;.
\eea
If $z=y$ or $b=0$, this can be solved to give 
\bea
\nonumber
x=x_{0}e^{-6 \epsilon \tau}\;,\\
\nonumber
y=y_{0}e^{2 \epsilon \tau}\;,\\
\nonumber
z=z_{0}e^{2 \epsilon \tau}\;,
\label{z}
\eea
where of course $y_0=z_0$ in the case $y=z$.

For $z \neq y$ and $b\neq0$, the perturbations behave like 
\bea
\nonumber
x=3 b^2(y_{0}-z_{0})(\eps e^{\eps \tau} +const)^2\;,\\
\nonumber
y=y_{0}e^{2 \epsilon \tau}\;,\\
\nonumber
z=z_{0}e^{2 \epsilon \tau}\;.
\eea

Notice that the Friedmann constraint (\ref{friedo}) translates 
into $0 \leq y-z\leq 1$, therefore in particular $y_0-z_0 \geq 0$. 

Since we are looking at small perturbations $0 < x(\tau) \ll 1$,  
we conclude that the constant of integration is positive for 
$\eps=-1$, whereas $const<0$ for $\eps=1$. This shows, 
that $M_{\epsilon}^0$ is a saddle for all values of $b$. It should be 
realised that for $b\neq 0$, the dynamics around the expanding 
model $M_{+}^0$ are reflecting the non\hs linearity of the system: 
$x$ is decreasing until $e^{\tau}=-const$, but then the system is repelled 
since $x$ increases for $-const<e^{\tau}<\infty$.
   
Table III below summarizes the dynamical character of the critical 
points with non\hs positive spatial curvature.

Note that if one of the eigenvalues in Table I vanishes, this just means 
that the lowest order terms vanish. This is very different to what happens 
in the case of systems of linear differential equations, where vanishing 
eigenvalues indicate lines of critical points. For a non\hs linear system 
to contain a line of critical points, we need small perturbations away 
from the critical point to be stationary in one direction. We found an example 
of that case in the analysis of  $m_\epsilon^0(\Omega_{\lambda})$ above. 
Since we are dealing with a non\hs linear system here, vanishing 
eigenvalues only indicate that it is not sufficient to look at the 
linearised equations, i.e. to study the eigenvalues and eigenvectors 
of the Jacobian. Instead, we carried out a perturbative analysis of the 
kind that we have presented above when analyzing $M_{\epsilon}^0$ and
$m_{\epsilon}^0(\Omega_{\lambda})$. Including the higher order 
contributions, we obtained the results presented in Tables III and VI.

\begin{table*}
\caption{\label{tab:table2} Eigenvalues and eigenvectors of the critical 
points with non-positive spatial curvature $^3R \leq 0$. We have 
defined $\psi=\sqrt{\frac{8}{b^2}-3}$. Notice that the first eigenvector 
of second row is $(-6\eps-\sqrt{3}b\sqrt{\frac{1-\Omega_{\lambda}}
{\gamma}},0,3\eps\Omega_{\lambda}(\Omega_{\lambda}-1))
\longrightarrow (1,0,0)$ as $\gamma \rightarrow 0$ for 
$\Omega_{\lambda}\neq 1$.}
\begin{ruledtabular}
\begin{tabular}{ccc}
Model  & Eigenvalues & Eigenvectors \\ \hline
$\mbox{m}_\epsilon^0(\Omega_{\lambda}\neq 1)~for~ b=0$ & $\epsilon(-6,-2,0)$ &
$(2,0,\Omega_{\lambda}(1-\Omega_{\lambda})),(0,1,-\Omega_{\lambda}),(0,0,1)$ \\
$\mbox{m}_\epsilon^0(\Omega_{\lambda}\neq 1)~for~b \neq 0$ & $(\infty,-2
\epsilon,0)$ & $(1,0,0),(0,-1,\Omega_{\lambda}),(0,0,1)$  \\
$\mbox{F}_\epsilon^2$ & $(6\epsilon+\sqrt{6}b,4\eps,-6\eps)$ & $(1,0,0),
(0,1,0),(0,0,1)$ \\
$\mbox{m}_\epsilon^2$ & $2\epsilon(3,5,3)$ & $ (1,0,0),(0,1,-1),(0,0,1)$
\footnote{We can use any two linearly independent vectors in the 
$\Omega_k=0$-plane as first and third eigenvectors; we have chosen the 
ones above for convenience.} \\
$\mbox{M}_\epsilon^2$ & $2\epsilon(3,-2,-5)$ & $(1,0,0),(0,1,0),(0,1,-1) $ \\
$\mbox{F}_{+}^{b^2/3}$ & $(\frac{b^2}{2}-3,b^2-2,-b^2)$ & $(1,0,0),
(\frac{b^2}{2}(\frac{b^2}{3}-2),\frac{b^2}{2}+1,0),(b^2(1-\frac{b^2}{6}),
0,3(\frac{b^2}{2}-1))$ \\
$\mbox{X}_{+}^{2/3}(b)$ & $(-1-\psi,-1+\psi,-2)$ & $(-\frac{2}{3} b^2,1
-\psi,0),(-\frac{2}{3}b^2,1+\psi,0),(-2,3(1-\frac{2}{b^2}),
-3(1-\frac{2}{b^2}))$ \\
\end{tabular}
\end{ruledtabular}
\end{table*}

\subsection{Models with Positive Spatial Curvature}
We now analyse the dynamical system (\ref{system.c}), which describes flat 
or closed models. In addition to the critical points corresponding to the 
flat models $F_{+}^{b^2/3}$, $F_{\epsilon}^2$, 
$m_{\eps}^0(\tilde{\Omega}_{\lambda})$ and $m_{\epsilon}^2$, we find 
a maximally non\hs general\hs relativistic Einstein universe\hs like model 
$E^{1/3}$ with $\gamma=1/3,~ k=1$ and $H=0$, which for $b=0$ degenerates 
into a line (hyperbola) of critical points $E(\tilde{\Omega}_{\lambda})$ 
with $\gamma\in [\textstyle{\frac{1}{3}},\textstyle{\frac{2}{3}}]$. 
For $0\leq b^2 \leq 2$, we also find the model $X_{+}^{2/3}(b)$, 
which has $\gamma =2/3$ and $Q= \mp b/ \sqrt{2}$. A summary of all 
these critical points, their coordinates in state space and their 
eigenvalues can be found in Table IV below. The corresponding 
eigenvectors are given in Table V.

In order to analyse the nature of $m_{\eps}^0(\tilde{\Omega}_{\lambda})$, 
we have to study the first order perturbations around the point, as shown 
in detail in the previous section. We find here, that the behaviour of the 
perturbed system agrees to leading order with the results obtained in 
section \ref{Positive}.

Notice that the Jacobian of the dynamical system around the equilibrium 
point corresponding to the static model $E^{1/3}$ is not well\hs defined 
for $b\neq0$ however we can extract the eigenvalues using a limiting 
procedure. The results are given in Table IV. Since the third eigenvalue 
vanishes, it is not sufficient to analyse the linearised equations at 
that point, unless we remain in the invariant subset $\tilde{\Omega}_{\lambda}
=1$, in which $E^{1/3}$ appears as a saddle with the two eigenvectors 
given in Table V. A study of the full perturbed equations around 
$E^{1/3}$ confirms, that this point is a saddle however it has a more 
complicated behaviour out of the $\tilde{\Omega}_{\lambda}$\hs plane; to 
second order, the system will be repelled in the $\tilde{\Omega}_{\lambda}$\hs 
direction in the expanding sector ($Q\geq0$) , while in the collapsing 
sector ($Q\leq0$), it is attracted in that direction.

The dynamical character of all the critical points with non\hs negative 
spatial curvature is summarized in Table VI.
\section{The structure of the state space} 
With the information we have obtained in the last section about 
the equilibrium points of the dynamical system, we can now analyse the 
structure of the state space. The whole state space is obtained by matching 
the dynamical systems (\ref{system.o}) and (\ref{system.c}). It consists 
of three pieces corresponding to the collapsing open models, the closed 
models and the expanding open models, matched together at the corresponding 
flat boundaries.

As mentioned above, the state space has the invariant 
subsets $\gamma=0,2$;~$\Omega_{k}=0 \Leftrightarrow Q=\pm1$;~ 
$\Omega_{\lambda},\tilde{\Omega}_{\lambda}=0$ and 
$\Omega_{k}+\Omega_{\lambda},\tilde{\Omega}_{\lambda}=1$. 

The bottom planes in Figures 1\hs 9 correspond to GR, while top planes 
represent the vacuum boundaries. 
 
Since the eigenvalues of the equilibrium points in general depend on the 
parameter $b$, the nature of these critical points varies with $b$. 
Furthermore, some of the critical points ($F_{+}^{b^2/3},~X_{+}^{2/3}$) 
will also move in state space as $b^2$ increases, since their 
coordinates are also functions of $b$. We also find critical points, 
which only appear for $b=0$ and then disappear, as we increase $b^2$. 
 
For certain values of $b$, some of the eigenvalues pass through zero. 
At these values of $b$, the state space is ``torn'' in that it undergoes 
topological changes. In contrast to the model that \cite{Campos} 
analysed, the appearance of such bifurcations in our model does not 
coincide with the appearance of lines of critical points. This a 
consequence of the fact, that in our inflationary model, the critical 
points are moving in state space. Bifurcations appear, when two of the 
equilibrium points merge, which occurs at $b^2=0,2,6$. These are the 
same parameter values as in GR \cite{Burd,Halliwell}. 
This was to be expected, since the only points moving in state space 
and causing these bifurcations, are the ones corresponding to the 
general relativistic models $F_{+}^{b^2/3},~X_{+}^{2/3}$. For this 
reason the occurrence of bifurcations is restricted to the GR\hs 
subspace. Therefore the dynamics of these bifurcations will only be 
discussed in section \ref{gr}.

Since the collapsing open sector remains unchanged with varying parameter 
value $b$, we do not include that sector in the different portraits of 
state space. The whole state space can be obtained by matching 
Fig. 1 to Fig. 2\hs Fig. 9.

\begin{table*}
\caption{\label{tab:table3} Dynamical character of the critical points 
with non\hs positive spatial curvature.}
\begin{ruledtabular}
\begin{tabular}{ccccccccc}
Model  & $b=0$ & $0<b^2<2$ & $b^2=2$ & $2<b^2<\frac{8}{3}$ & 
$b^2=\frac{8}{3}$ & $\frac{8}{3}<b^2<6$ & $b^2=6$ & $b^2>6$ \\ \hline
$\mbox{m}_{+}^0(\Omega_{\lambda})$ & line of sinks & line of saddles & 
line of saddles & line of saddles & line of saddles & line of saddles & 
line of saddles & line of saddles \\ 
$\mbox{m}_{-}^0(\Omega_{\lambda})$ & line of sources & line of sources & 
line of sources & line of sources & line of sources & line of sources & 
line of sources & line of  sources \\
$\mbox{M}_\epsilon^0$ & saddle & saddle & saddle & saddle & saddle & 
saddle & saddle & saddle \\ 
$\mbox{M}_\epsilon^2$ & saddle & saddle & saddle & saddle & saddle & 
saddle & saddle & saddle \\ 
$\mbox{F}_{+}^2$ & saddle & saddle & saddle & saddle & saddle & 
saddle & saddle & saddle \\
$\mbox{F}_{-}^2$ & saddle & saddle & saddle & saddle & saddle & 
saddle & saddle & saddle \\
$\mbox{m}_{+}^2$ & source & source & source & source & source & 
source & source & source \\
$\mbox{m}_{-}^2$ & sink & sink & sink & sink & sink & sink & sink & sink \\
$\mbox{F}_{+}^{b^2/3}$ &  line of sinks & sink & saddle 
\footnote{Notice that this point is an attractor for all open or flat 
models, while it is a repeller for all closed models.}  & saddle & 
saddle & saddle & saddle & -\\
$\mbox{X}_{+}^{2/3}(b)$ & $-$ & - & saddle $^a$ & sink & sink & spiral sink & spiral sink & 
spiral sink\\
\end{tabular}
\end{ruledtabular}
\end{table*}

\subsection{The GR\hs subspace} \label{gr} 
As discussed above, the subset $\Omega_{\lambda},\tilde{\Omega}_{\lambda}=0$ 
is the invariant submanifold of the full state space corresponding to GR. 
In this section, we will discuss the stability of the general 
relativistic equilibrium points within the GR\hs subspace. In the next 
section, we will discuss the brane\hs world\hs modifications of these 
general relativistic results due to the additional degrees of 
freedom $\Omega_{\lambda},~\tilde{\Omega}_{\lambda}$, and also the 
additional non\hs general relativistic equilibrium points.

It is worth mentioning that although the GR\hs submanifold has been  
discussed in detail by Burd and Barrow \cite{Burd} and Halliwell 
\cite{Halliwell}, their analysis is somewhat incomplete, since 
the state space they considered was non\hs compact. 
The variables they chose to describe the inflationary dynamics of 
homogeneous isotropic models were $\phi ', \alpha '$. Here $\alpha$ 
relates to the scale factor $a(\tau)$ by $a(\tau)=e^{\alpha(\tau)}$, 
and the potential has been absorbed into the time derivative by 
rescaling the time variable $t \rightarrow \tau$ by 
$\frac{d}{d\tau}=~'~=V^{-1/2}\frac{d}{dt}$. We find that these 
coordinates relate to our expansion normalized variables via the following 
transformations:
\bea
\nonumber
(\alpha')^2=\frac{2}{3}\frac{1}{(2-\gamma)(1-\Omega_{k})}\;,\\
\nonumber
(\phi')^2=\frac{2\gamma}{2-\gamma}\;,
\eea
for $\gamma\neq2,~\Omega_{k}\neq1$. We can see that
\begin{equation}
\nonumber
\alpha ' \rightarrow \infty ~\Leftrightarrow ~\Omega_{k}\rightarrow 1 
~\Leftrightarrow~ \Omega_{\rho} \rightarrow 0
\end{equation}
and
\begin{equation}
\nonumber
\phi'\rightarrow \infty~ \Leftrightarrow ~\gamma \rightarrow 2\;.
\end{equation}
This means, that we have compactified the state space by mapping 
$\phi '\in [0,\infty] \longrightarrow \gamma \in [0,2]$ and  
$\alpha '\in [0,\infty] \longrightarrow \Omega_{k} \in [0,1]$. In this way
we have extended the work that has been done on inflationary models with 
exponential potentials in the general relativistic context. We find 
that $X_{+}^{2/3},~F_{+}^{b^2/3}$ correspond to the critical points I, II 
in \cite{Burd}. In addition, we obtain the new critical points 
$F_{\eps}^2$, $M_{\eps}^0$ and $M_{\eps}^2$ corresponding to a FLRW 
universe with stiff matter and a Milne universe with $\gamma=0$ or stiff 
matter $\gamma=2$ respectively.

Notice that we also find the new equilibrium point $F_{\eps}^0$. The reason
is that we are using $\gamma$ to describe the dynamics of the scalar field 
$\phi$, which essentially is a function of $\dot{\phi}^2$ instead 
of $\dot{\phi}$. Therefore our $\dot{\gamma}$\hs equation is homogeneous 
in $\dot{\phi}$, whereas the dynamical equation for $\dot{\phi}$ is 
inhomogeneous in $\dot{\phi}$.

It might seem surprising, that we find a critical point at 
$\dot{\phi}=0$, whereas there is no equilibrium point at $\dot{\phi}=0$ 
in \cite{Burd}. The reason for this is, that we are describing different 
physical quantities; although the equation of state does not change 
as $\dot{\phi} \rightarrow 0~(\dot{ \gamma}=0)$, $\dot{\phi}$ does change 
as $\dot{\phi} \rightarrow 0 ~,(\ddot{\phi},\phi''\neq0)$.

The collapsing FLRW model with stiff matter found in this analysis is 
of particular interest, since for all values of $b$, $F_{-}^2$ is the 
future attractor for all collapsing open and some of the closed models 
within the GR\hs subspace.  $F_{-}^0$ is the past attractor for the whole 
collapsing open and parts of the closed sector for all $b$.

Notice that the dynamics of the collapsing open sector does not depend on 
the parameter value $b$. The dynamics of this sector is constrained by the 
fact, that for all models in this sector, the flat FLRW model with 
constant energy density is the past attractor, and the flat FLRW model 
with stiff matter is the future attractor (see FIG 1).

Having said that, we will now discuss the more complicated 
dynamics of the closed and the expanding open models for the different 
ranges of the parameter value $b$.

\begin{table*}
\caption{\label{tab:table4} Coordinates and eigenvalues of the 
critical points with non\hs negative spatial curvature. We have 
defined $\chi=\sqrt{\frac{8-3b^2}{2}}, 
\phi=\sqrt{\tilde{\Omega}_{\lambda}^2+3\tilde{\Omega}_{\lambda}+1}$}
\begin{ruledtabular}
\begin{tabular}{ccc}
Model  & Coordinates   & Eigenvalues \\ \hline
$\mbox{m}_\epsilon(\tilde{\Omega}_{\lambda}\neq 1)~for~ b=0$ & 
$(0,1,\tilde{\Omega}_{\lambda})$ & $-2\epsilon(3,1,0)$ \\
$\mbox{m}_\epsilon^0(\tilde{\Omega}_{\lambda}\neq 1)~for~ b \neq 0$ & 
$(0,1,\tilde{\Omega}_{\lambda})$ & $(\infty,-2\epsilon,0)$ \\
$\mbox{F}_\epsilon^2$ & $(2,\epsilon,0)$ & $(6\epsilon+\sqrt{6}b, 
4\eps,-6\eps)$ \\
$\mbox{m}_\epsilon^2$ & $(2,\epsilon,1)$ & $2\epsilon(3,5,3)$ \\
$\mbox{F}_{+}^{b^2/3}$ & $(\frac{b^2}{3},1,0)$ & $(\frac{b^2}{2}-3, 
b^2-2,-b^2)$\\
$\mbox{X}_{+}^{2/3}(b)$ & $(\frac{2}{3},-\frac{b}{\sqrt{2}},0)$ & 
$(\frac{b}{\sqrt{2}}-\chi,\frac{b}{\sqrt{2}}+\chi,\sqrt{2}b)$\\
$\mbox{E}^{1/3}~$ & $(\frac{1}{3},0,1)$ &
$(\sqrt{5},-\sqrt{5},0)$\\
$\mbox{E}(\tilde{\Omega}_{\lambda})~for~ b=0$ & $(\frac{2}{3(1+
\tilde{\Omega}_{\lambda})},0,\tilde{\Omega}_{\lambda})$ &
$(\frac{2}{1+\tilde{\Omega}_{\lambda}}\phi,-\frac{2}{1+
\tilde{\Omega}_{\lambda}}\phi,0)$ \\ 
\end{tabular}
\end{ruledtabular}
\end{table*}

At $b=0$, we find a bifurcation of the state space, since for this value 
of $b$, the equilibrium points $E^{1/3}$ and $X_{+}^{2/3}(b)$, as well 
as $F_{+}^0$ and $F_{+}^{b^2/3}$ coincide. All expanding open models 
are attracted to $F_{+}^0$. All closed models will evolve into 
either $F_{-}^2$ or $F_{+}^0$, depending on the initial conditions. 
At $\gamma=2/3$, the Einstein universe appears as an unstable saddle 
point (see FIG 2). 

As $b^2$ increases, the Einstein universe disappears. Instead, we find 
the saddle point $X_{+}^{2/3}(b)$, which moves in $Q$\hs direction as 
$b^2$ is increases, remaining a saddle until it reaches the flat 
sector $Q=1$ at $b^2=2$. There it merges with expanding FLRW model 
$F_{+}^{b^2/3}$. For $0<b^2<2$, $F_{+}^{b^2/3}$ is a sink moving along 
the $\gamma$\hs axis. All open and some of the closed models are  
attracted to this solution (see FIG 3).

At $b^2=2$, the models $X_{+}^{2/3}(b)$ and $F_{+}^{b^2/3}$ merge, which 
causes a bifurcation of the state space. The nature of the two merging 
critical points is significantly changed at this value of $b$. The flat 
solution $X_{+}^{2/3}(b=-\sqrt{2})=F_{+}^{2/3}$ is an attractor for all 
open or flat models, but a repeller for all closed models (see FIG 4). 
This behaviour has been observed by \cite{Burd}. At this value of $b^2$, 
the two merging critical points swap their nature: for all 
$b^2\geq 2$, $X_{+}^{2/3}(b)$ will be a sink, while $F_{+}^{b^2/3}$ will 
be a saddle (see FIGS 5\hs 9).

As $b$ is further increased ($2<b^2<6$), $F_{+}^{b^2/3}$ moves 
further along the $\gamma$\hs axis. It is now a saddle for all models. 
All flat models are attracted, whereas all open or closed models are 
repelled. $X_{+}^{2/3}(b)$ has now entered the open sector and moves 
further to the $\Omega_{k}=1$ boundary. $X_{+}^{2/3}(b)$ is a node sink 
for all $2<b^2<8/3$ and a spiral sink for $8/3<b^2<\infty$. For 
all $b^2 \geq 2, ~F_{-}^2$ is the future attractor for all closed models 
and $X_{+}^{2/3}(b)$ the future attractor for all expanding open models 
(see FIGS 5\hs 7).   

At $b^2=6$, we find another bifurcation of the state space. The two 
points  $F_{+}^{b^2/3}$ and $F_{+}^2$ merge. This turns $F_{+}^2$ from 
a source into a saddle (see FIG. 8). For $b^2\geq 6$, all open or closed 
models are still repelled from $F_{+}^2$, but all flat models are now 
attracted. In the limit $b^2 \rightarrow \infty$, the future attractor 
for the open sector, $X_{+}^{2/3}(b)$ approaches the vacuum 
solution $\Omega_{\rho}\rightarrow 0$ with $\gamma=2/3$ (see FIG 9).  

\begin{table*}
\caption{\label{tab:table5} Eigenvalues and eigenvectors of the critical 
points with non-negative spatial curvature $^3R \geq 0$. We have 
defined $\chi=\sqrt{\frac{8-3b^2}{2}} $, 
$\phi=\sqrt{\tilde{\Omega}_{\lambda}^2+3\tilde{\Omega}_{\lambda}+1}$. 
Notice that the first eigenvector of second row is 
$(-6\eps-\sqrt{3}b\sqrt{\frac{1-\tilde{\Omega}_{\lambda}}{\gamma}},
0,3\tilde{\Omega}_{\lambda}(\tilde{\Omega}_{\lambda}-1))
\longrightarrow (1,0,0)$ as $\gamma \rightarrow 0$ for $\tilde{\Omega}_{\lambda}\neq 1$.}
\begin{ruledtabular}
\begin{tabular}{ccc}
Model  & Eigenvalues & Eigenvectors \\ \hline
$\mbox{m}_\epsilon^0(\tilde{\Omega}_{\lambda}\neq 1)~for~b=0$ 
& $\epsilon(-6,-2,0)$ & $(2,0,\tilde{\Omega}_{\lambda}(1
-\tilde{\Omega}_{\lambda})),(0,1,0),(0,0,1)$ \\
$\mbox{m}_\epsilon^0(\tilde{\Omega}_{\lambda}\neq 1)~for~b \neq 0$ 
& $(\infty,-2\epsilon,0)$ & $(1,0,0),(0,1,0),(0,0,1)$ \\
$\mbox{F}_\epsilon^2$ & $(6\epsilon+\sqrt{6}b,4\eps,-6\eps)$ 
& $(1,0,0),(0,1,0),(0,0,1)$ \\
$\mbox{m}_\epsilon^2$ & $2\epsilon(3,5,3)$ 
& $ (1,0,0),(0,1,0),(0,0,1)$ \footnote{We can use any two linearly 
independent vectors in the $\Omega_k=0$-plane as first and third 
eigenvectors; we have chosen the ones above for convenience.} \\
$\mbox{F}_{+}^{b^2/3}$ & $(\frac{b^2}{2}-3,b^2-2,-b^2)$ & $(1,0,0),
(b^2(\frac{b^2}{3}-2),\frac{b^2}{2}+1,0),(b^2(1-\frac{b^2}{6}),
0,3(\frac{b^2}{2}-1))$ \\
$\mbox{X}_{+}^{2/3}(b)$ & $(\frac{b}{\sqrt{2}}-\chi,\frac{b}{\sqrt{2}}
+\chi,\sqrt{2}b)$ & $ (\frac{8}{3},\frac{b}{\sqrt{2}}
+\chi,0),(\frac{8}{3},\frac{b}{\sqrt{2}}-\chi,0),(2,\frac{3}{2\sqrt{2}}
b(1-\frac{b^2}{2}),3(1-\frac{b^2}{2}))$\\
$\mbox{E}^{1/3}~$ & $(\sqrt{5},-\sqrt{5},0)$ & $(\sqrt{5},-3,0),
(\sqrt{5},3,0),- $\\
$\mbox{E}(\tilde{\Omega}_{\lambda})~for~b=0$ & $(\frac{2}{1
+\tilde{\Omega}_{\lambda}}\phi,-\frac{2}{1+\tilde{\Omega}_{\lambda}}\phi,0)$
& $(-\frac{4+6\tilde{\Omega}_{\lambda}}{3(1+\tilde{\Omega}_{\lambda})},
\phi,\tilde{\Omega}_{\lambda}(\tilde{\Omega}_{\lambda}-1)),(\frac{4
+6\tilde{\Omega}_{\lambda}}{3(1+\tilde{\Omega}_{\lambda})},\phi,
-\tilde{\Omega}_{\lambda}(\tilde{\Omega}_{\lambda}-1)),(1,0,
-\frac{3}{2}(1+\tilde{\Omega}_{\lambda})^2) $ \\ 
\end{tabular}
\end{ruledtabular}

\end{table*}

\subsection{Higher dimensional effects} 
The question we will now discuss is how the behaviour of the dynamical 
system describing GR is changed within the brane\hs world context. 
In particular, do the additional degree of freedom 
$\Omega_{\lambda},~\tilde{\Omega}_{\lambda}$ change the stability 
of the general relativistic models, and are there new non\hs general 
relativistic stable equilibrium points?

We will first answer the second question. In addition to the general 
relativistic points, we have found the non\hs general\hs relativistic 
equilibrium points $m_{\eps}^2$ and the lines $m_{\eps}^0(\Omega_{\lambda})$. 
We can solve the Friedmann equation at these points to determine their 
behaviour in detail. At $m_{\eps}^0$, the energy density is 
constant $\rho(t)=\rho_0$, and we can easily integrate the Friedmann 
equation to find 
\begin{equation}
a(t)\sim e^{\epsilon\sqrt{\tilde{\Lambda}/3}~t}\;,
\end{equation}
where $\tilde{\Lambda}=\kappa^2\rho_0(1+{\textstyle{\frac{\rho_0}{2\lambda}}})$ behaves like a modified cosmological constant.

For $m_{+}^2$, the scale factor is given by 
\begin{equation}
a(t)=(t-t_{BB})^{1/6}(t+t_{BB})^{1/6}\;. 
\end{equation}
where $t_{BB}=1/\sqrt{6\kappa^2\lambda}$ is the Big Bang time.
As explained in detail in \cite{Campos}, this model corresponds to 
the Bin\'{e}truy\hs Deffayet\hs Langlois (BDL) solution 
\cite{Binetruy} with scale factor $a(t)=(t-t_{BB})^{1/6}$. The 
solution for $m_{-}^2$ is given by time reversal and $t_{BB}$ should 
now be identified as the Big Crunch time.

At $b=0$, we also observe, that the general\hs relativistic Einstein universe 
has non\hs general\hs relativistic analogues. We find a whole line of 
Einstein universe\hs like static equilibrium points 
$E(\tilde{\Omega}_{\lambda})$ extending in the $\tilde{\Omega}_{\lambda}$ 
direction. For $b^2>0$, the line collapses to the non\hs general\hs 
relativistic Einstein saddle $E^{1/3}$ in the $\tilde{\Omega}_{\lambda}=1$ 
subset, and the general relativistic model $X_{+}^{2/3}(b)$.    

We now study the dynamical character of these non\hs general\hs relativistic 
equilibrium points. We find that in the full state space, $m_{-}^2$ instead 
of $F_{-}^2$ is the future attractor for all collapsing open models and some 
of the closed models. This means, that within the brane\hs world context, FLRW 
with stiff matter changes from a stable solution into a unstable saddle.  
$F_{-}^0$ remains a past attractor for the collapsing open and parts of the 
closed sector, but now there is a whole line $m_{-}^0(\Omega_{\lambda})$ 
of sources extending in $\tilde{\Omega}_{\lambda}$ direction. 
The same applies to the corresponding expanding models: the sink/saddle 
$F_{+}^0$ is a one\hs element subset of the line of sinks/saddles 
$m_{+}^0(\Omega_{\lambda})$. This means, that for $b=0$, the future 
attractor of the expanding open and some of the closed models is not 
necessarily $F_{+}^0$, but instead any of the points $m_{+}^0(\Omega_{\lambda})$ depending on the initial conditions. For $b^2>0$, 
$m_{+}^0(\Omega_{\lambda})$ including $F_{+}^0$ turns into a 
line of saddles.

For $b>0$ the models $m_{+}^0(\Omega_{\lambda})$ represent high energy 
inflationary models with exponential potentials (which are too steep to
inflate in GR) \cite{Copeland}. The fact that they are all unstable 
(saddle points) reflects the fact that {\it steep inflation} ends 
naturally, since as the energy drops below the brane tension the 
condition for inflation no longer holds.

Finally let us consider whether the stability of the most interesting 
equilibrium points $F_{+}^{b^2/3}$ and $X_{+}^{2/3}$ is changed by the 
higher\hs dimensional degrees of freedom. From Tables V, VI, we can see that 
the third eigenvalues (corresponding to eigenvectors pointing out of the 
GR\hs plane) are negative for all $b\neq0$. That means, if $F_{+}^{b^2/3}$, 
$X_{+}^{2/3}$ was a sink in GR, it will remain a sink in the brane\hs world 
scenario.   

\begin{table*}
\caption{\label{tab:table6} Dynamical character of the critical points 
with non\hs negative spatial curvature.}
\begin{ruledtabular}
\begin{tabular}{ccccccc}
Model  & $b=0$ & $0<b^2<2$ & $b^2=2$ & $2<b^2<6$ & $b^2=6$ & $b^2>6$ \\ \hline
$\mbox{m}_{+}^0(\tilde{\Omega}_{\lambda})$ & line of sinks & line of saddles 
& line of saddles & line of saddles & line of saddles & line of saddles \\ 
$\mbox{m}_{-}^0(\tilde{\Omega}_{\lambda})$ & line of sources & line of 
sources & line of sources & line of sources & line of sources & line of 
sources \\
$\mbox{F}_{+}^2$ & saddle & saddle & saddle & saddle & saddle & saddle \\
$\mbox{F}_{-}^2$ & saddle & saddle & saddle & saddle & saddle & saddle \\
$\mbox{m}_{+}^2$ & source & source & source & source & source & source \\
$\mbox{m}_{-}^2$ & sink & sink & sink & sink & sink & sink \\
$\mbox{F}_{+}^{b^2/3}$ & line of sinks & sink & saddle 
\footnote{Notice that this point is an attractor for all open or flat 
models, while it is a repeller for all closed models. } & saddle & saddle & -\\
$\mbox{X}_{+}^{2/3}(b)$ & line of saddles & saddle & saddle $^a$ & - & - & - \\
$\mbox{E}^{1/3}$ & saddle & saddle & saddle & saddle & saddle & saddle \\ 
$\mbox{E}(\tilde{\Omega}_{\lambda})~for~ b=0$ & line of saddles & - & - & - 
& - & - \\ 
\end{tabular}
\end{ruledtabular}
\end{table*}

\section{Discussion and Conclusion}
In this paper we extend recent work by Campos and Sopuerta \cite{Campos}
to the case where the matter is described by a dynamical scalar field $\phi$ 
with an exponential potential. By using expansion normalised variables 
which compactifies the state space we built on earlier results due to 
Burd and Barrow \cite{Burd} and Halliwell \cite{Halliwell} for the case 
of GR and explored the effects induced by higher dimensions in the 
brane\hs world scenario. 

As in \cite{Campos} we obtain the equilibrium point corresponding 
to the BDL model ($\mbox{m}_\pm$) \cite{Binetruy} which dominates 
the dynamics at high energies (near the Big Bang and Big Crunch), where 
the extra\hs dimension effects become dominant supporting the 
claim that this solution is a generic feature of the state space of 
more general cosmological models in the brane\hs world scenario.

We emphasise again that here, unlike in the analysis by 
Campos and Sopuerta \cite{Campos}, $\gamma$ is a dynamical variable. 
Fixing $\gamma$ to be a constant corresponds to looking at 
the $\gamma=const.$ slices of the full state space. This obviously 
only makes sense for the invariant sets $\gamma=0$ and $\gamma=2$. 
The important point is, that even if we want to analyse the dynamical 
character of the de Sitter and Milne models in the $\gamma=0$\hs plane, 
we have to bear in mind the dynamical character of $\gamma$. Unlike in 
\cite{Campos}, we have to study perturbations away from the plane. This makes
 these  models much more interesting in the presence of an exponential 
potential, since the dynamics are not reduced to the plane $\gamma=0$. 
In fact, unless $b=0$, we find that the plane $\gamma=0$ is unstable. Small
 perturbations out of that plane will in general be enhanced, 
i.e. even for initial conditions with negligible $\gamma$, the system will 
in general evolve towards $\gamma=2/3$ or $\gamma=2$. Notice that orbits 
confined to the $\gamma=0$ plane evolve towards the expanding de Sitter 
models $m^0_+(\Omega_{\lambda})$  or the contracting Milne universe 
$M^0_{-}$ for all values of $b$.

Finally we note that we did not find any new bifurcations in this simple 
brane\hs world scenario because we consider only the case ${\cal U}=0$. In 
the next paper in this series \cite{paper2} we will analyse both the 
effects of the non\hs local energy density ${\cal U}$ on the FLRW 
brane\hs world dynamics and look at homogeneous and anisotropic models
with an exponential potential.  
 
{\bf Acknowledgments:}
We would like to thank Roy Maartens and Varun Sahni for 
very useful discussions during the Cape Town Cosmology Meeting in 
July 2001. We also thank Toni Campos and Carlos Souperta for comments
and suggestions for future work. This work has been funded by the 
National Research Foundation (SA) and a UCT international postgraduate 
scholarship.



\begin{figure*}
\includegraphics{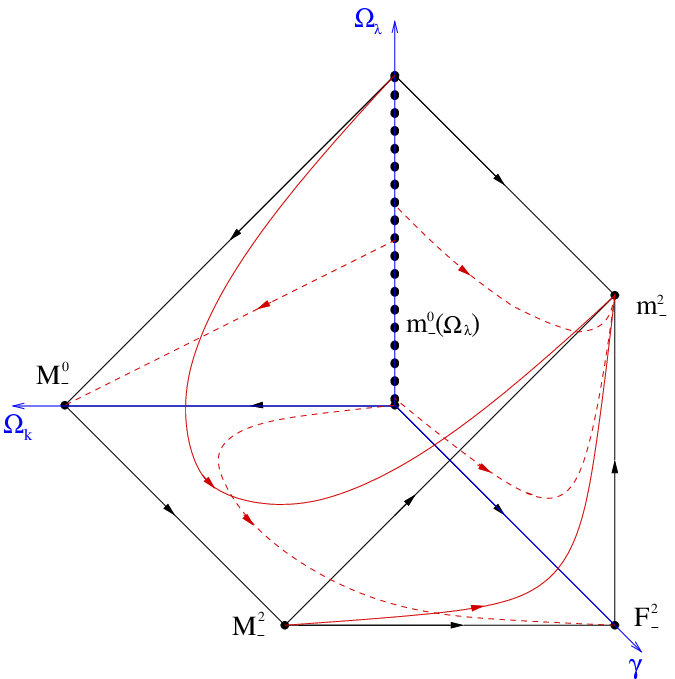}
\caption{\label{fig:Fig1} State space for the collapsing open FLRW models, $\eps=-1,~^3R \leq 0$. The bottom plane $\Omega_{\lambda}=0$ corresponds to general relativity. The top surface $\Omega_k+\Omega_{\lambda}=1$ corresponds to vacuum $\Omega_{\rho}=0$. The equilibrium points $M_{-}^0,~M_{-}^2,~F_{-}^2,~m_{-}^2,~m_{-}^0(\Omega_{\lambda})$ describe the Milne universe with $\gamma=0$ / stiff matter, the flat FLRW model with stiff matter, the non-general-relativistic BDL model with stiff matter and a line of non-general-relativistic models with constant energy density (including the flat FLRW model with $\gamma=0$). 
We are only drawing the trajectories in the invariant planes, from which the whole dynamics can be deduced. 
The structure of this part of the whole state space does not change with the parameter value $b$. The full state space can be obtained by matching this section to FIG. 2 - FIG. 9.}
\end{figure*}

\begin{figure*}
\includegraphics{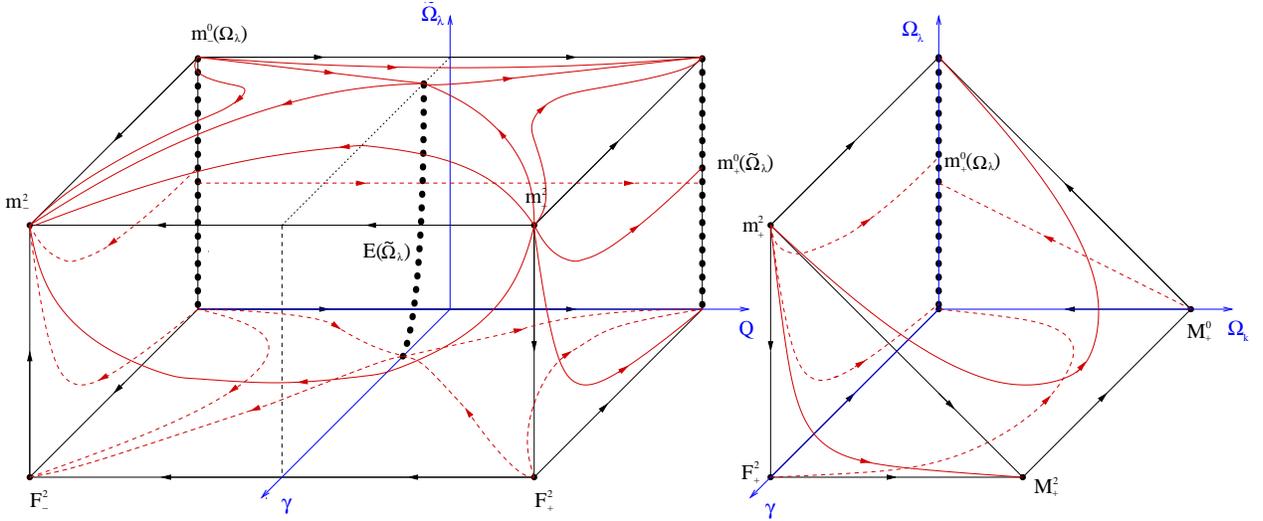}
\caption{\label{fig:Fig2} State space for the FLRW models with non-negative 
spatial curvature $^3R \geq 0,~\eps= \pm 1$ (on the left) and the expanding 
FLRW models with non-positive spatial curvature, $^3R \leq 0,~\eps=1$ 
(on the right) for $b=0$ (a bifurcation). In the left part of the figure, 
which describes the closed models, the plane $Q=0$ differentiates between the expanding 
sector $Q \geq 0,~ \eps=1$ and the collapsing sector $Q \leq 0,~\eps=-1$. 
As in FIG. 1, the bottom plane corresponds to GR, whereas the top surfaces 
represent the vacuum solutions. We only give the trajectories on 
the invariant planes, from which the whole dynamics can be deduced.
The critical points $F_{\eps}^2,~m_{\eps}^2,~m_{\eps}^0(\tilde{\Omega}
_{\lambda}),~E(\tilde{\Omega}_{\lambda}) $ correspond to the flat FLRW 
model with stiff matter, the non-general-relativistic BDL model with 
stiff matter, a line of non-general relativistic model with constant 
energy density (including the flat FLRW model with $\gamma=0$) and a 
line of static Einstein universes with $1/3 \leq \gamma \leq 2/3$.}
\end{figure*}

\begin{figure*}
\includegraphics{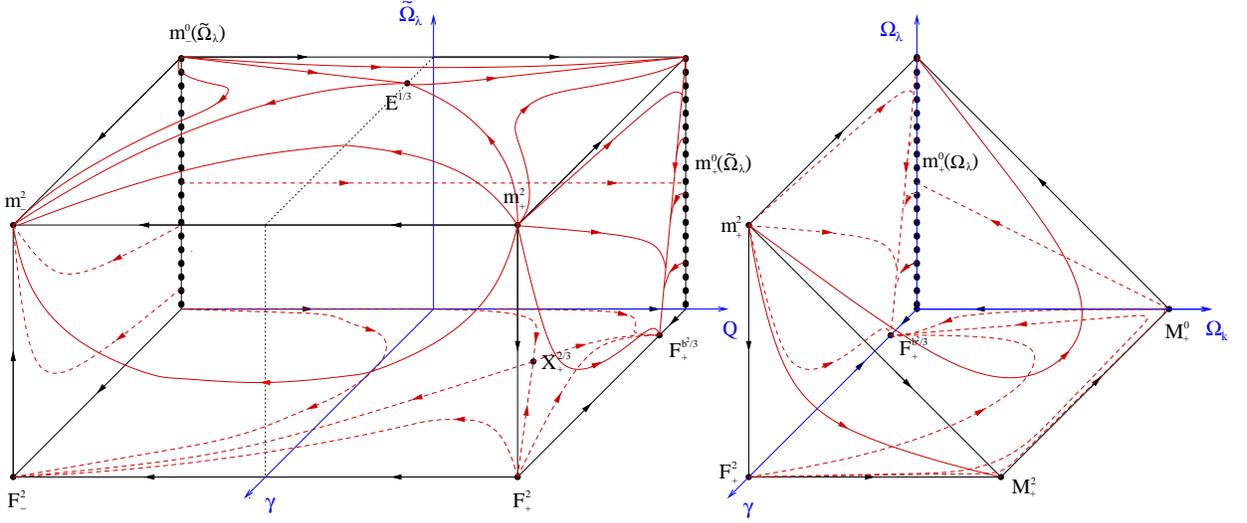}
\caption{\label{fig:Fig3} State space for the FLRW models with $0<b^2<2$. 
The line of Einstein universes in the closed sector of the state space 
has collapsed into the non-general-relativistic Einstein model $E^{1/3}$ 
and the general-relativistic model $X_{+}^{2/3}$, which is expanding and 
moving towards the expanding flat subspace $Q=1$, ($\eps=1$) as $b^2$ is 
increasing. The equilibrium point $F_{+}^{b^2/3}$ corresponds to a flat 
FLRW model with $\gamma = b^2/3$. See the captions of FIG. 1, FIG. 2 for 
more details.}
\end{figure*}

\begin{figure*}
\includegraphics{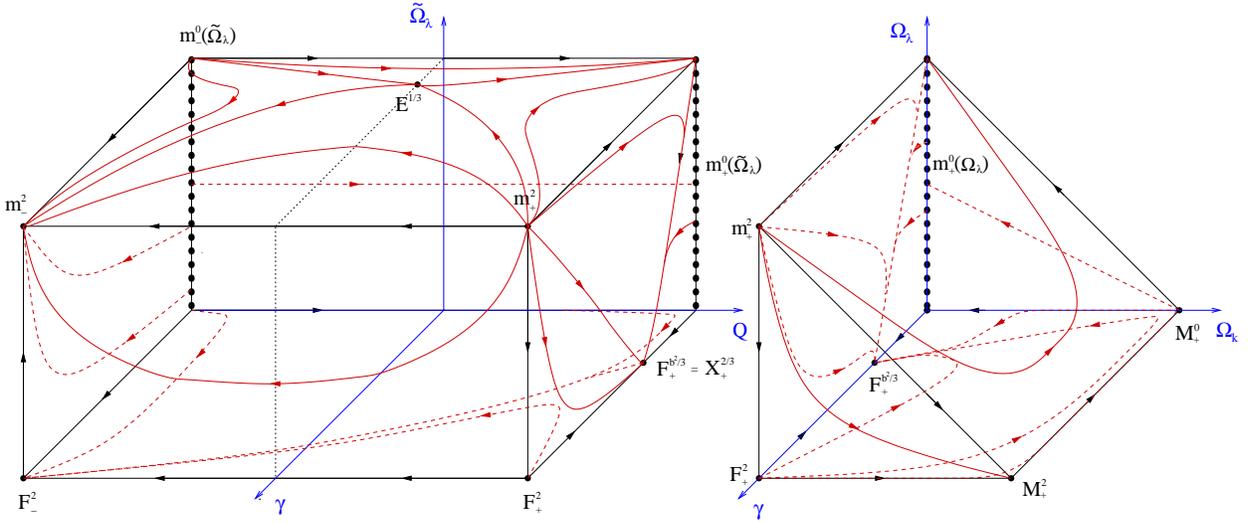}
\caption{\label{fig:Fig4} State space for the FLRW models with $b^2=2$ (a bifurcation). The equilibrium point $X_{+}^{2/3}$ has reached the flat subspace, where it merges with $F_{+}^{b^2/3}$. This causes the bifurcation; the nature of the two critical points will be swapped as they are moving on (see FIG. 5 - 9). All open and flat models are attracted to this point $X_{+}^{2/3}=F_{+}^{b^2/3}$, whereas all closed models are repelled. See the captions of FIG. 1, FIG. 2 for more details.}
\end{figure*}

\begin{figure*}
\includegraphics{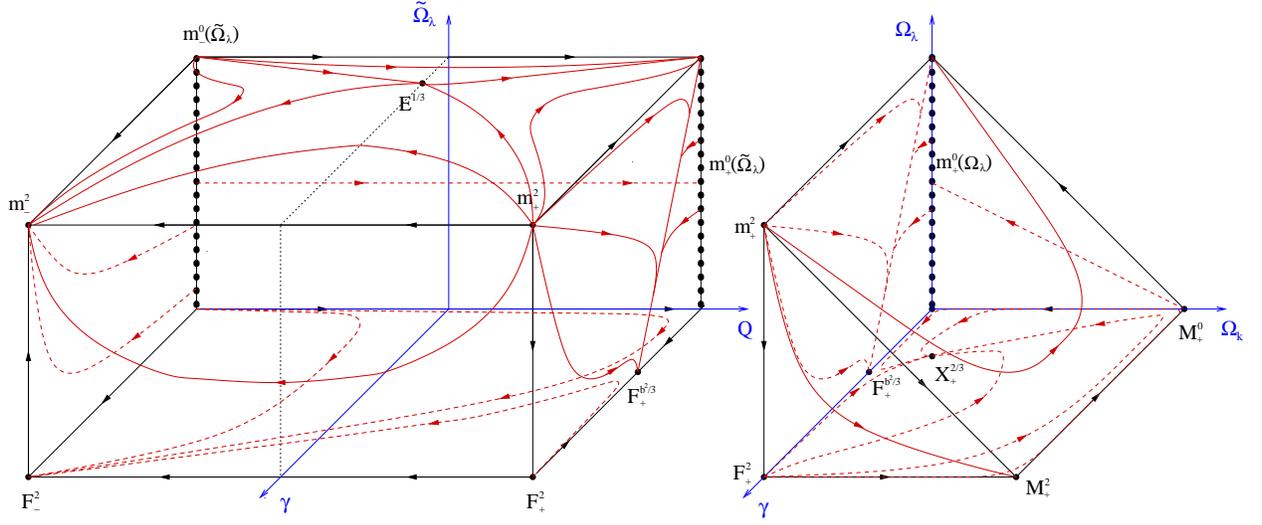}
\caption{\label{fig:Fig5} State space for the FLRW models with $2<b^2<8/3$. $X_{+}^{2/3}$ has entered the open sector and turned into a node sink. $F_{+}^{b^2/3}$ is now an unstable saddle for all models. See the captions of FIG. 1, FIG. 2 for more details.}
\end{figure*}

\begin{figure*}
\includegraphics{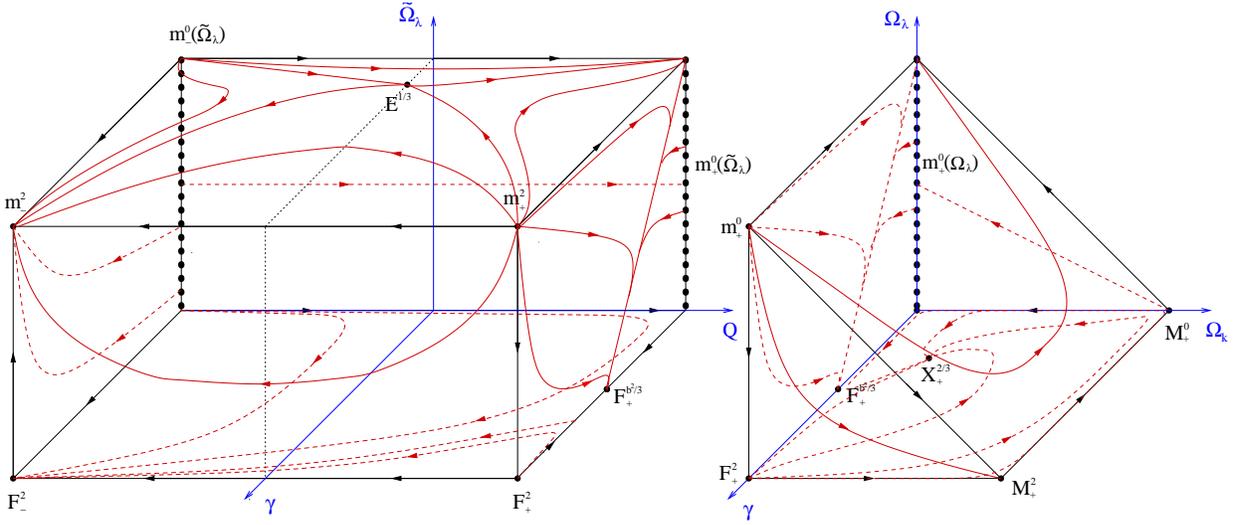}
\caption{\label{fig:Fig6} State space for the FLRW models with $b^2=8/3$. The node sink $X_{+}^{2/3}$ is turning into a spiral sink; the structure of the state space has not changed. See the captions of FIG. 1, FIG. 2 for more details.}
\end{figure*}

\begin{figure*}
\includegraphics{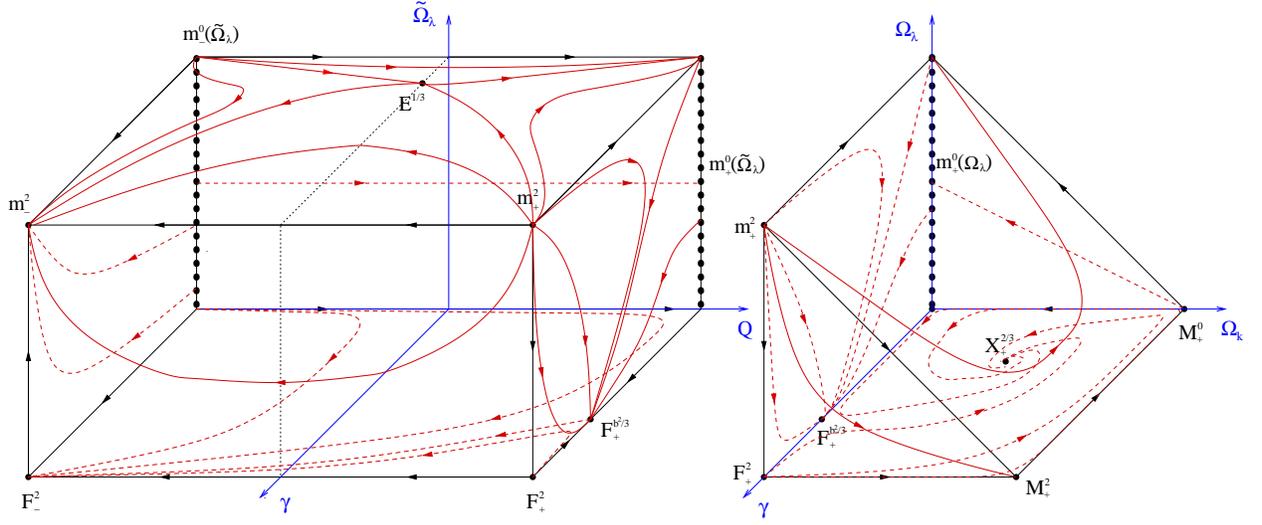}
\caption{\label{fig:Fig7} State space for the FLRW models with $8/3<b^2<6$. $X_{+}^{2/3}$ is now a spiral sink; the structure of the state space is essentially the same as for $2<b^2 \leq 8/3$. See the captions of FIG. 1, FIG. 2 for more details.}
\end{figure*}

\begin{figure*}
\includegraphics{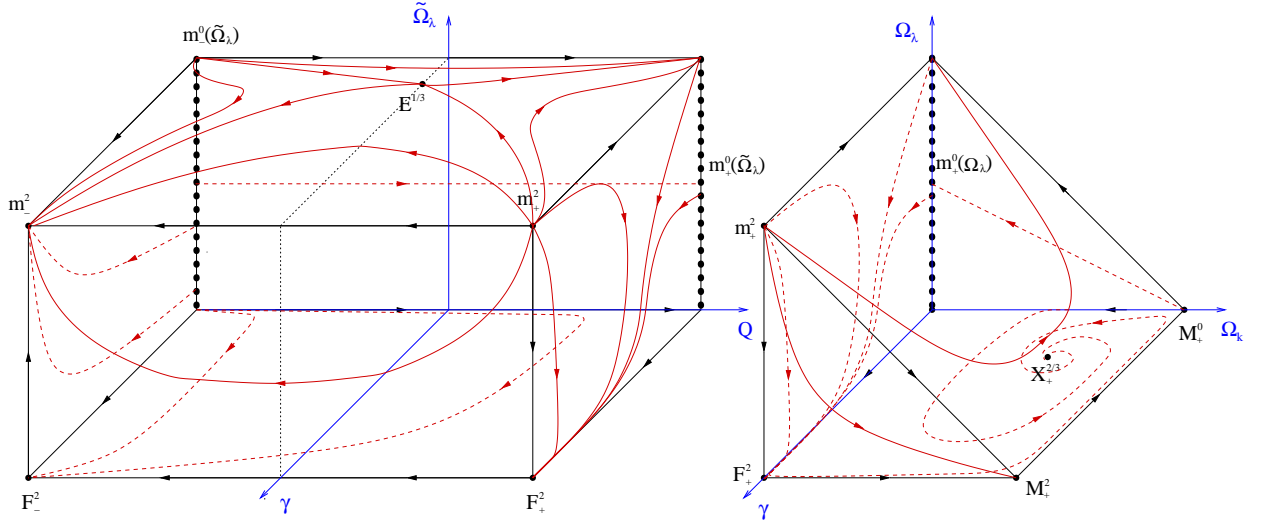}
\caption{\label{fig:Fig8} State space for the FLRW models with $b^2=6$ (a bifurcation). The equilibrium point $F_{+}^{b^2/3}$ has merged with the equilibrium point $F_{+}^2$, which causes the bifurcation. The topology of the state space is changing at this value of $b$, since the point $F_{+}^2$ is turning from an attractor in $\gamma$-direction into a repeller in that direction (compare to FIG. 7, FIG. 9). See the caption of FIG. 1, FIG. 2 for more details.}
\end{figure*}

\begin{figure*}
\includegraphics{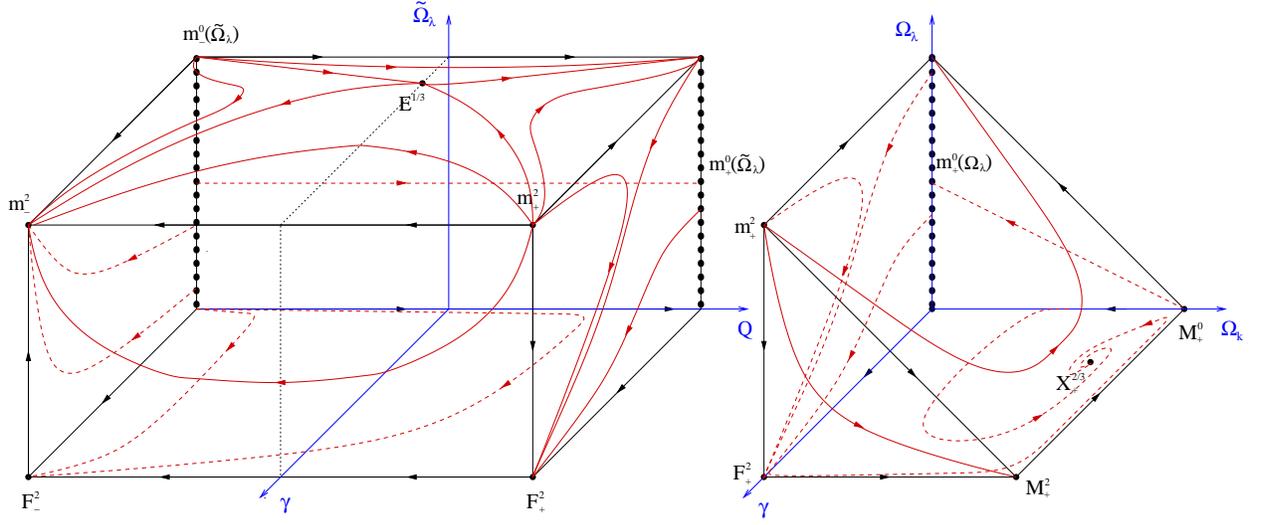}
\caption{\label{fig:Fig9} State space for the FLRW models with $b^2>6$. The model  $F_{+}^{b^2/3}$ has moved out of state space, while $X_{+}^{2/3}$ is approaching the open vacuum boundary of the GR-subspace. See the captions of 
FIG. 1, FIG. 2 for more details.}
\end{figure*}

\end{document}